\documentclass[reprint,superscriptaddress,longbibliography]{revtex4-1}
\usepackage{amsmath,graphicx}
\usepackage{siunitx}
\usepackage{dirtytalk}
\usepackage{comment}
\usepackage{color}

\begin{document}

\title{Measurement of variations in gas refractive index with $10^{-9}$ resolution\\using laser speckle}

\author{Morgan Facchin} \email{mf225@st-andrews.ac.uk}
\affiliation{SUPA, School of Physics and Astronomy, University of St Andrews, North Haugh, St Andrews KY16 9SS, UK}

\author{Graham D. Bruce}
\affiliation{SUPA, School of Physics and Astronomy, University of St Andrews, North Haugh, St Andrews KY16 9SS, UK}

\author{Kishan Dholakia}
\affiliation{SUPA, School of Physics and Astronomy, University of St Andrews, North Haugh, St Andrews KY16 9SS, UK}
\affiliation{Department of Physics, College of Science, Yonsei University, Seoul 03722, South Korea}
\affiliation{School of Biological Sciences, The University of Adelaide, Adelaide, South Australia, Australia}

\begin{abstract}

Highly-resolved determination of refractive index is vital in fields ranging from biosensing through to laser range-finding. Laser speckle is known to be a sensitive probe of the properties of the light and the environment, but to date speckle-based refractive index measurements have been restricted to $10^{-6}$ resolution. In this work we identify a strategy to optimise the sensitivity of speckle to refractive index changes, namely by maximising the width of the distribution of optical path lengths in the medium. We show that this can be realised experimentally by encapsulating the medium of interest within an integrating sphere. We demonstrate that variations of the refractive index of air as small as $4.5\times10^{-9}$ can be resolved with an uncertainty of $7\times10^{-10}$. This is an improvement of three orders of magnitude when compared to previous speckle-based methods.
\end{abstract}
\maketitle

\section{Introduction}

Refractive index is a parameter of importance across most areas of optical measurement. It can be used in cell biology to investigate particular cell metabolic activities or as a probe of other biophysical quantities \cite{Liu16}, and in chemical sensing to measure concentrations of liquids \cite{Yunus88}.  Interferometric measurements of length and displacement in gaseous environments \cite{Kruger16} are limited in their accuracy by uncertainties in the refractive index of the environment. High-precision measurements of refractive index have even been proposed as a route to a more accurate definition of the Pascal \cite{Silander20}. Small changes in refractive index can have major implications: Infection can cause the refractive index of red blood cells to change at the $10^{-3}$ level \cite{Park08}; biosensors measure cell secretion dynamics and protein concentrations by tracking refractive index changes at the $10^{-5}$ level \cite{Li17,Conteduca21}.
In optical tweezers experiments exploring the motion of RNA polymerase during transcription \cite{Abbondanzieri05}, the effect of air currents (which typically modulate the refractive index at the order of $10^{-7}$ \cite{Fang99,Khelifa98}) caused sufficient position instability of the optical trapping and measurement beams to mask the angstrom-level motion, even in a sealed environment. There are many methods to measure refractive index using lasers, including hollow-core \cite{Chen20}, photonic crystal \cite{Quan15,Zhang19} or evanescent optical fibre refractometers \cite{Li17b} (fibre-based devices have been recently reviewed in \cite{Urrutia19}), and metasurfaced-based refractometers \cite{Conteduca21}. The most sensitive measurements of refractive index in the literature are variants of double-channel Fabry-Perot cavities \cite{Khelifa98}, with which refractive index uncertainties of $10^{-12}$ have recently been demonstrated \cite{Silander20}.

Laser speckle, formed when a coherent light field interacts with a disordered medium, is a powerful probe of changes of the laser or the medium itself, and is therefore an attractive tool to harness in sensing applications. The first application of speckle in refractometry was presented half a century ago \cite{Kopf72,Debrus72}, and most subsequent work has adopted a similar scheme. A laser beam impinges on a random phase screen to produce a speckle field, which then traverses a medium under investigation, and the changes in the speckle pattern can be used to quantify changes in the refractive index. Speckle has been applied to measurements of the refractive indices of air \cite{Lapsien97}, glass \cite{Guo18}, and liquids \cite{Trivedi19}. Recently, by immersing up to three consecutive planar diffusers inside a medium of interest, Tran et al used the resultant speckle to measure refractive index with a resolution on the order of $10^{-6}$ \cite{Tran20}.

How might one further optimise the sensitivity of a speckle refractometer? A speckle pattern is the result of the interference of many different wave paths. When light propagates in a medium of refractive index $n$, the phase acquired on a given path of length $z$ is $n\,k\,z$, with $k$ the wavenumber. A change in refractive index therefore applies a phase shift $\Delta n \,k\,z$ on that path. Now any change occurring in the speckle pattern results from relative phase changes between paths. It follows that what maximises the sensitivity of the speckle pattern is not path length itself, as intuition might suggest, but the \emph{spread} in path lengths within the medium of interest. In a simple illustrative case where we consider only two given paths, their relative phase changes by $k \Delta n \Delta z$, where we see that the maximal effect is obtained for a maximal path length difference. This strategy is consistent with the work of Tran et al \cite{Tran20} where the speckle sensitivity is greatly increased by the use of successive planar diffusers inside the medium, increasing the number of paths and their length differences. However, that approach is still limited by the globally paraxial geometry of the diffusion. 

In this work, to further increase the spread in path lengths and enable more finely-resolved measurements of refractive index, we take inspiration from recent progress in speckle-based measurements of the wavelength of monochromatic light \cite{chakrabarti2015speckle,Mazilu14,hanson2015speckle,Metzger17,odonnellhigh,davila2020single,gupta19}. In that context, we have recently analytically shown the advantages of using an integrating sphere to generate the speckle \cite{Facchin_model}. The integrating sphere creates a particularly broad path length distribution which offers orders of magnitude improvement in sensitivity compared with other speckle-based techniques. 

To quantify the change in the speckle due to a refractive index change, we develop a method based on the evaluation of the speckle similarity. This speckle similarity is then used to extract the refractive index change, based on an explicit relation existing between the two, that we derive analytically and verify experimentally. By varying the pressure of air inside an integrating sphere, we resolve changes in refractive index as low as $4.5\times10^{-9}$ with an uncertainty of $7\times10^{-10}$. Our method surpasses all previous speckle refractometers and has the potential to match the existing state-of-the-art measurements of small changes in the refractive index of gases.

\section{similarity profile}
The quantitative tool that we use in this work is the speckle similarity, or correlation, which quantifies the change occurring in a speckle pattern. It is defined as 

\begin{equation} \label{eq:correl}
S= \Big\langle  \Big( \frac{I_{i} - \langle I_{i}\rangle}{\sigma_{I}} \Big)\Big( \frac{I_{i}'-\langle I_{i}'\rangle}{\sigma_{I'}} \Big) \Big\rangle ,
\end{equation}
with $I$ and $I'$ two speckle images (before and after a change), $\sigma_{I}$ and $\sigma_{I'}$ their respective standard deviation, and the angular brackets denoting averaging over the image. This gives a value of~1 for identical images and decreases towards 0 as they diverge from one another. 

It was shown in \cite{Facchin_model} that for two speckle patterns taken before and after some generic transformation, the similarity is given by 

\begin{equation} \label{Sgeneral}
S =   \frac{ 1 }
{\big (1-\frac{\sigma^2}{2\ln{\rho}} \big)^2+\big (\frac{\mu}{\ln{\rho}} \big)^2     },
\end{equation}

\noindent where $\mu$ and $\sigma^2$ are respectively the mean and variance of the phase shift induced by the transformation on a single pass of light through the sphere, with $\rho$ the sphere's surface reflectivity. 

Let us apply this to the case of a change in refractive index of the medium filling the sphere. The phase light acquires on a given path of length $z$ is $n\,k\,z$, with $n$ the refractive index and $k$ the wavenumber. After a refractive index change, the phase changes by an amount $\Delta n \,k\,z$. It follows that the average phase change on a single path is $\mu=\Delta n \,k\,\overline{z}$, with $\overline{z}$ the average chord length in the sphere, which is given by geometry to be $4R/3$ \cite{Fry:06,berengut1972random,sidiropoulos2014n}. Likewise, the standard deviation of chord length is $\sqrt{2}R/3$ \cite{berengut1972random,sidiropoulos2014n}. This gives 

\begin{equation} \label{stat}
\mu=\frac{4}{3}\Delta n \,k R \quad \quad \quad \sigma=\frac{\sqrt{2}}{3}\Delta n \,kR.
\end{equation}

Inserting this in (\ref{Sgeneral}), it can be shown that the $\mu$ term dominates, which leaves us with a Lorentzian profile:

\begin{equation} \label{eq:lorentzian_refrac}
S=  \frac{1}{1+\left(\frac{\Delta n}{ \Delta n_0}\right)^2 },
\end{equation} 
with $\Delta n_0=3\lambda \left | \ln\rho\right | /(8\pi R) $, which also corresponds to the HWHM of the Lorentzian. For modest parameters such as $R=1$ cm, $\rho=$ 0.9, and $\lambda=$ 780 nm, this gives $\Delta n_0 = 10^{-6}$.

This similarity profile can be used in a very simple way to determine the refractive index difference $\Delta n$ between two given times, by taking the reciprocal function 

\begin{equation} \label{eq:inversion}
\Delta n = \Delta n_0 \sqrt{1/S-1},
\end{equation} 
with $S$ the similarity between the two corresponding speckles. As we will show, refractive index changes much smaller than $\Delta n_{0}$ can be resolved - the ultimate resolution of the method depends on the detector noise in recording the speckle.

\section{Experimental implementation}
\begin{figure}[htbp!]
\centering\includegraphics[width=\linewidth]{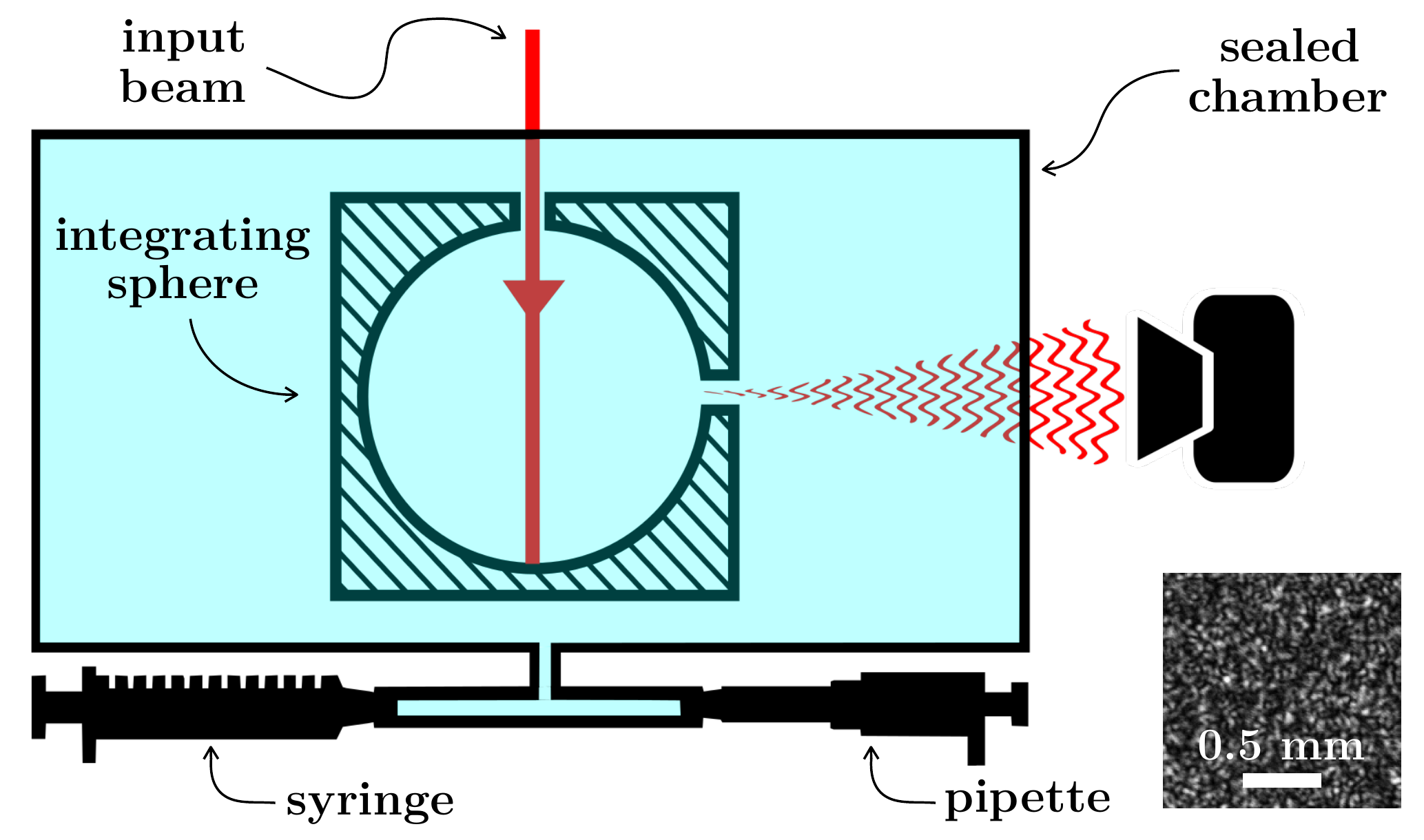}
\caption{Experimental setup. An integrating sphere is placed in a sealed chamber, which light can enter and exit via two glass windows. The air inside the chamber is slightly compressed by pushing a syringe (for the verification of (\ref{eq:lorentzian_refrac})) or a pipette (for the measurement of small variations). The resulting change in refractive index of the air induces a change in the speckle pattern which is recorded on a camera. An example of the resultant speckle pattern is shown. }
\label{fig:setup}
\end{figure}

In this section we experimentally verify relation (\ref{eq:lorentzian_refrac}), with the setup described in Fig. \ref{fig:setup}. A laser beam of wavelength 780~nm, 10 mW power, and a coherence length of a few kilometres (Toptica DLPro) is injected into an integrating sphere, and the resulting speckle pattern in collected on a CMOS camera (Mikrotron MotionBLITZ EoSens mini2). We use a 1.25 cm radius sphere, carved into a 3 cm edge aluminium cube and coated with Spectraflect to make a Lambertian and highly reflective surface. The light enters and escapes the sphere through two 3~mm diameter holes. The sphere is placed in a $2490\pm 50$~ml stainless steel chamber which is hermetically sealed using CF flanges and copper gaskets. 

\begin{figure}[htbp!]
\centering\includegraphics[width=\linewidth]{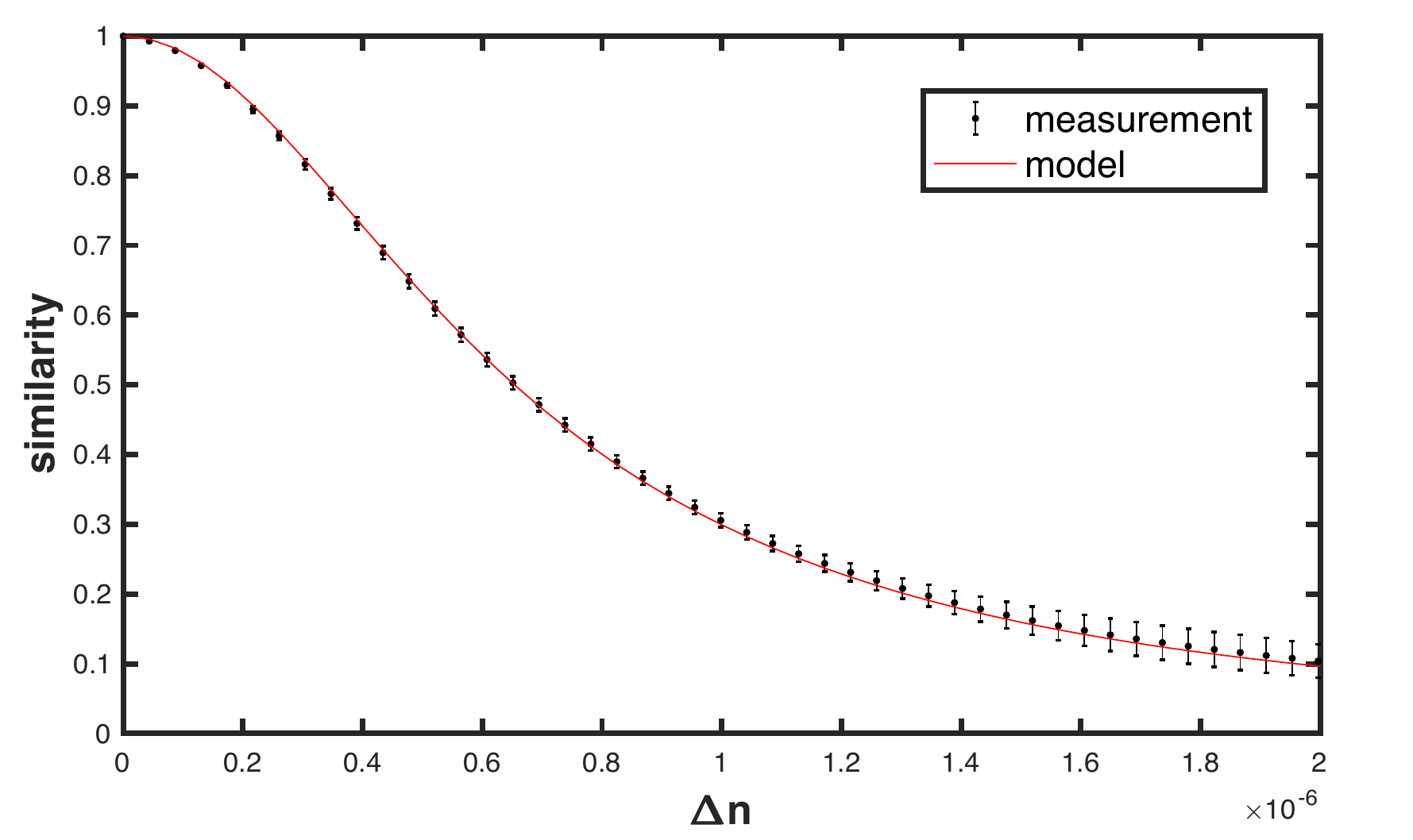}
\caption{Speckle similarity as a function of refractive index change, experimental (black dots) and Lorentzian profile predicted by model (red line), fitted for a reflectivity $\rho=0.916$. The centre and span of the error bars respectively give the mean and standard deviation of a set of curves extracted from the data set. The HWHM is $6.5\times10^{-7}$.}
\label{fig:similrefrac}
\end{figure}

The refractive index variations are obtained by slightly compressing the air inside the chamber using a $100$ ml syringe connected to the chamber via a needle valve. The syringe is compressed at a constant rate of 4.0~\si{ml.s^{-1}} using a motorised translating stage while the changing speckle is recorded. From this we extract the similarity profile as a function of refractive index change shown in Fig.~\ref{fig:similrefrac}. The value of the refractive index change is inferred from the volume change by the following. The fact that the chamber is sealed implies $\Delta n/n'=\Delta \rho_{\text{air}}/\rho_{\text{air}}\approx -\Delta V/V$, with $n=1+n'$ the refractive index of the air inside the chamber, $\rho_{air}$ its density, and $V$ the chamber's volume, assuming $n'\propto\rho_{\text{air}}$ (Gladstone-Dale law \cite{gladstone1863xiv}) and $\Delta V  \ll V$. It follows that the refractive index change is given by $\Delta n=-n'\Delta V / V$, with $n'=2.7\times 10^{-4}$ for our measured values of $\lambda=780$ nm, 20°C, and $100.5$ kPa \cite{ciddor1996refractive}. The main source of uncertainty is the volume of the chamber (2\%). In Fig. \ref{fig:similrefrac} we also display the uncertainty of the profile, given by the standard deviation of a set of curves extracted from the data set by using different reference images. By fitting the resulting profile using (\ref{eq:lorentzian_refrac}) with the reflectivity as a free parameter, we find best agreement for $\rho=0.916\pm0.002$, which is consistent with our previous estimations \cite{Facchin_model}. The HWHM is found to be $\Delta n_{0} = $ $6.5\times10^{-7}$, corresponding to a volume variation of only 6.0 ml, or a fractional volume change of 0.24\%. This can be done very easily by hand without feeling any pressure resistance, which offers a very hands-on demonstration of speckles' sensitivity.

\section{Measurement of small refractive index changes}
In this section we describe the measurement of refractive index variations much smaller than $\Delta n_0$. For such variations, using directly (\ref{eq:inversion}) is not ideal, as for $\Delta n\approx 0$ we have $dS/d\Delta n\approx 0$. This problem would be solved if we could look at small variations of the similarity around its point of maximal slope, which occurs at $\Delta n=\Delta n_0$, instead of around $\Delta n= 0$. This in fact can be done by purposely applying an initial refractive index variation of $\Delta n_0$ prior to the measurement. In this way the similarity, taken between a speckle before and after the initial $\Delta n_0$ leap, varies around a value of 0.5 with maximal sensitivity. In our setup, this initial variation could be applied by changing the volume of the chamber by 6.0 ml. 
However, a simpler way is to make use of the equivalence that exists between refractive index variation and wavelength variation. Specifically, the phase shift resulting from a wavelength change on a path of length $z$ is $n \, \Delta k \, z$, which is of the same form as what we found for a refractive index change ($\Delta n \,k \,z$). As both phase shifts are proportional to $z$, equating them on one path equates them on all paths, and the two effects are physically equivalent when $\Delta n = -\Delta \lambda/\lambda$ (with $n\approx1$). This means that the same change in a speckle pattern occurs after a refractive index change $\Delta n$ or after a wavelength change $\Delta \lambda=-\lambda \Delta n$. We can therefore bring the similarity to its point of maximal slope by applying an appropriate wavelength offset, in our case equal to 0.5 pm. 

\begin{figure}[htbp!]
\centering\includegraphics[width=\linewidth]{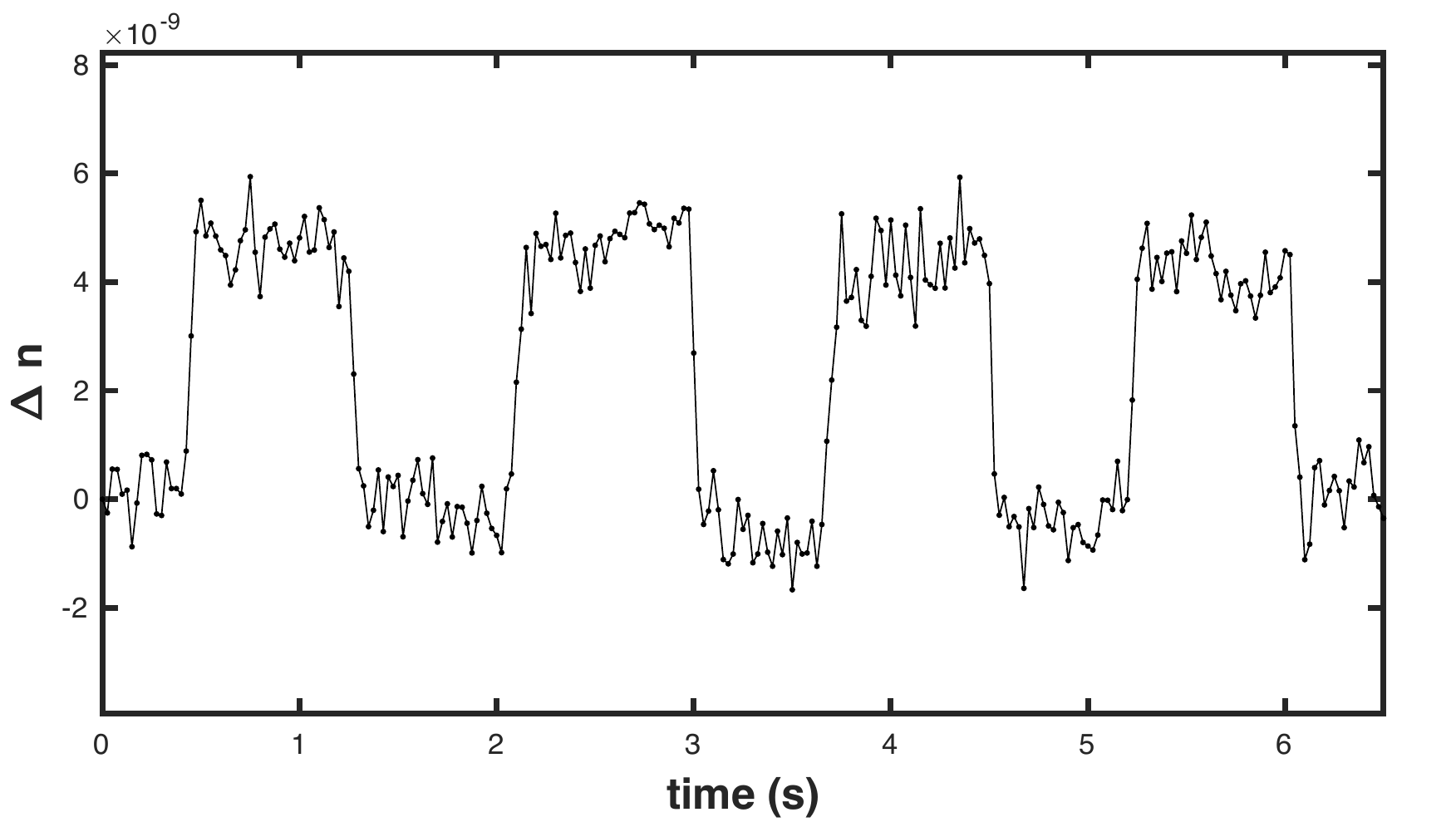}
\caption{Measurement of small periodic steps in refractive index, applied by changing the volume of the chamber by 40 $\mu$l using a pipette, corresponding to a fractional volume change of $(1.61 \pm 0.03)\times10^{-5}$. We find a step amplitude of $(4.5\pm0.7)\times10^{-9}$, in accord with the expected value of $(4.3\pm0.1)\times10^{-9}$. }
\label{fig:steps}
\end{figure}

We proceed in the following way. A reference speckle is first recorded at an initial wavelength. The wavelength is then offset by about 0.5 pm (this does not need to be precise). Thereafter, small refractive index changes are applied using a pipette instead of a syringe, which can apply much smaller volume changes. We press and release the pipette in a square wave manner with a period of about a second, with a 40 $\mu$l volume load. This corresponds to a fractional volume change of $(1.61 \pm 0.03)\times10^{-5}$, from which we infer an expected refractive index change of $(4.3\pm0.1)\times10^{-9}$. We compute the similarity between the reference speckle and the speckles undergoing change, which is then converted to refractive index difference using (\ref{eq:inversion}). The resulting curve is shown in Fig.~\ref{fig:steps}. The first value of the time series is subtracted, so that what is displayed is the refractive index variation applied by the pipette. We find steps of amplitude $(4.5\pm0.7)\times10^{-9}$, which is consistent with the expected value. 

The main source of noise is camera noise which propagates to the estimation of the similarity. The uncertainty on the value of the similarity is a function of the image size, which we empirically find to be $\delta S \approx 0.1/\sqrt{N}$ in our illumination conditions, with $N$ the number of pixels. For our image size, 200$\times$200, this gives $\delta S\approx5\times10^{-4}$. As the slope of the similarity curve at $\Delta n=\Delta n_0$ is $1/(2\Delta n_0)$, it follows that the corresponding uncertainty on the refractive index variation $\delta \Delta n$ is $2\Delta n_0\delta S$. With our parameters, this gives $\delta \Delta n=7\times10^{-10}$.

\section{Heating effect and compensation}
\begin{figure}[htbp!]
\centering\includegraphics[width=\linewidth]{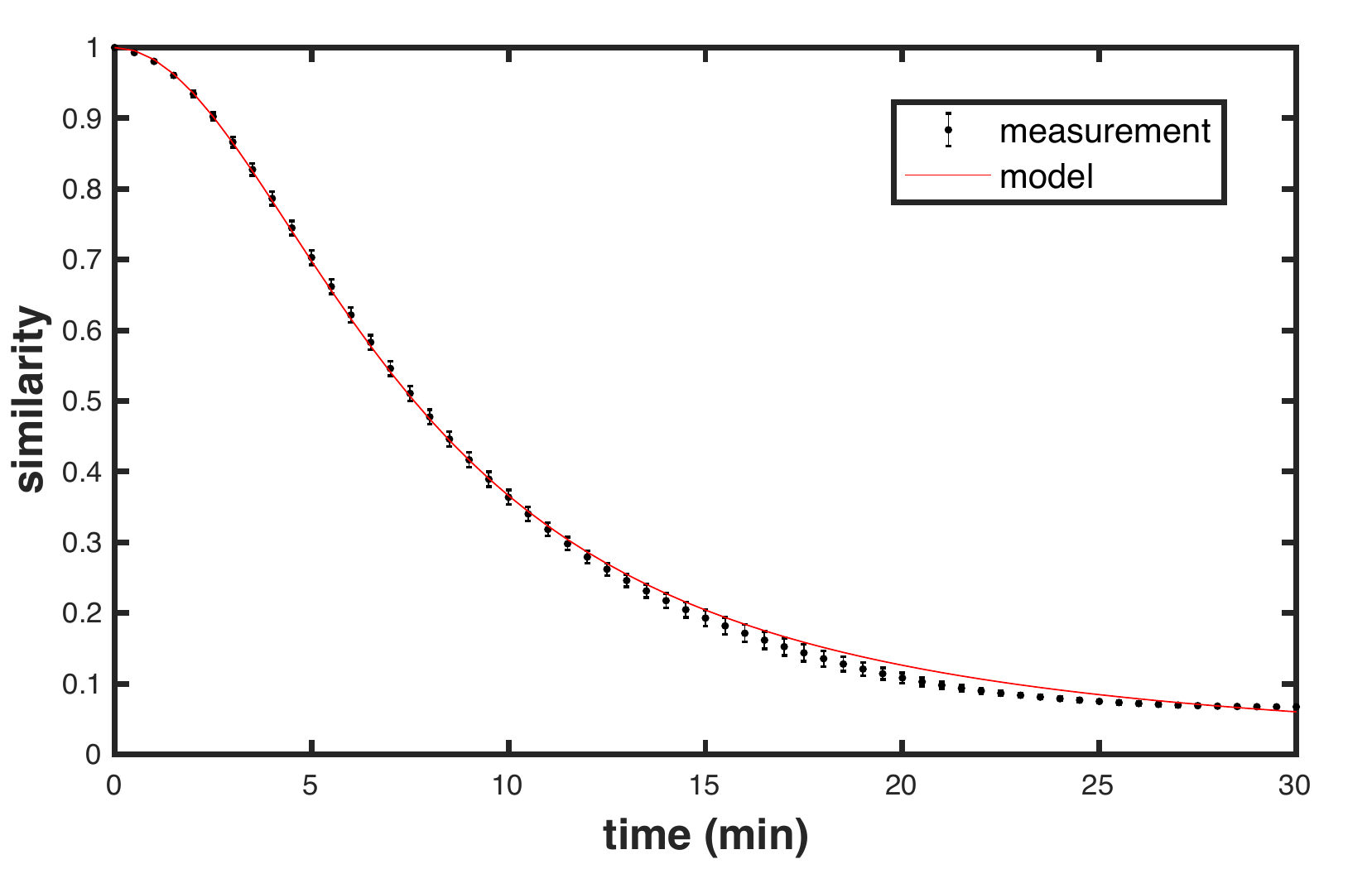}
\caption{Intrinsic change in the speckle pattern over time when no transformation is applied, due to heating from the laser. Black: similarity between the speckle at a reference time and subsequent times. Red: Lorentzian fit with a HWHM of 7.6 min. We can infer from this that the radius of the sphere increases by 1.1 nm every minute.   }
\label{fig:drift}
\end{figure}

When the system described in Fig. \ref{fig:setup} is left on its own without applying any transformation, we still observe a slow change of the speckle pattern over time. When quantifying this change by the similarity, we obtain the curve shown in Fig. \ref{fig:drift}, which (surprisingly) is also well fitted by a Lorentzian profile, with a HWHM of 7.6 min. We make the hypothesis that this time evolution comes from heating due to the input light. Inside the sphere, the diffusion of light is such that the surface power density is uniform across the inner surface. Therefore, in a steady state, nearly all the input power is absorbed uniformly on the inner surface (the power of the escaping light is negligible). The heat can then either be conducted to the material, or to the air inside the sphere. 

In the case of the material, an order of magnitude estimation shows that the heat diffusion time is very short (of the order of seconds), so that we can assume the increase in temperature to be uniform throughout the material. This temperature increase, in turns, leads to an isotropic thermal expansion of the sphere. The effect of such an expansion can be found analytically: the phase acquired by the field on a given path of length $z$ being $nkz$, and as an isotropic expansion increases all lengths by a factor $\Delta R/R$ (with $\Delta R$ the variation of radius resulting from the expansion), the resulting average phase shift on a single pass is $4k\Delta R/3$ (with $n\approx1$). Assuming in a first approximation that the heat remains stored in the sphere's material, we have $\Delta R \propto t$, which inserted in (\ref{Sgeneral}) leads indeed to a Lorentzian profile in time. One could expect that the increase in temperature of the sphere's material in turn induces a heat flux from the sphere to the surrounding air, leading to thermal equilibrium and stopping the thermal expansion. However we observed that the HWHM of the Lorentzian does not change significantly in over 80 minutes of measurements, meaning that no thermal equilibrium is reached in that time. This indicates that the approximation of the heat remaining in the sphere's material is valid at least in that time scale. Given that after a time equal to the HWHM we have the relation $4k\Delta R/3=\ln{\rho}$, we can infer that the radius of the sphere increases by 8.1 nm every 7.6 minutes by thermal expansion, which corresponds to a speed of 1.1 $\text{nm.min}^{-1}$ or 18 $\text{pm.s}^{-1}$.

We also see that as the phase shift due to thermal expansion has the same form as that of a refractive index change, we can also draw an equivalence between those two effects, given by the substitution $\Delta n \Leftrightarrow \Delta R/R $. This means that we can apply a refractive index change to compensate the thermal expansion. Denoting $t_0$ the observed HWHM of the Lorentzian profile in time, it can be shown that the volume rate that must be applied to compensate the expansion is $\dot{V}=-3V\ln{\rho}/(4 n' kR t_0)$. For $t_0=7.6$ min as found in Fig. \ref{fig:drift}, and with the same parameters as above, we have $\dot{V}=0.79$ \si{ml.min^{-1}}. The volume of the chamber was continuously increased at this rate in the measurement of Fig. \ref{fig:steps} to compensate for the thermal expansion. Here, contrary to the previous measurement, a change in volume is simpler to implement than a change in wavelength. That is because here the applied rate of change is small and constant in time, which is simpler to implement using a syringe on a commercial syringe pump (WPI AL2000).

On the other hand, we expect that the effect of the fraction of heat that goes to the air cannot be compensated. 
Indeed, in that case, the temperature increase has no reason to be uniform throughout the volume inside the sphere. If the effect if not uniform, the phase acquired by the field on a given path is no longer a simple function of its length and therefore cannot be compensated by a volume change. The relative amounts of heat that go into the material and into the air must depend on the properties of the material, in particular its thermal diffusivity. For that reason we think that most of the heat is conducted to the material in our case, as our sphere is made of aluminium, which has a high thermal diffusivity.

\begin{figure}[htbp!]
\centering\includegraphics[width=\linewidth]{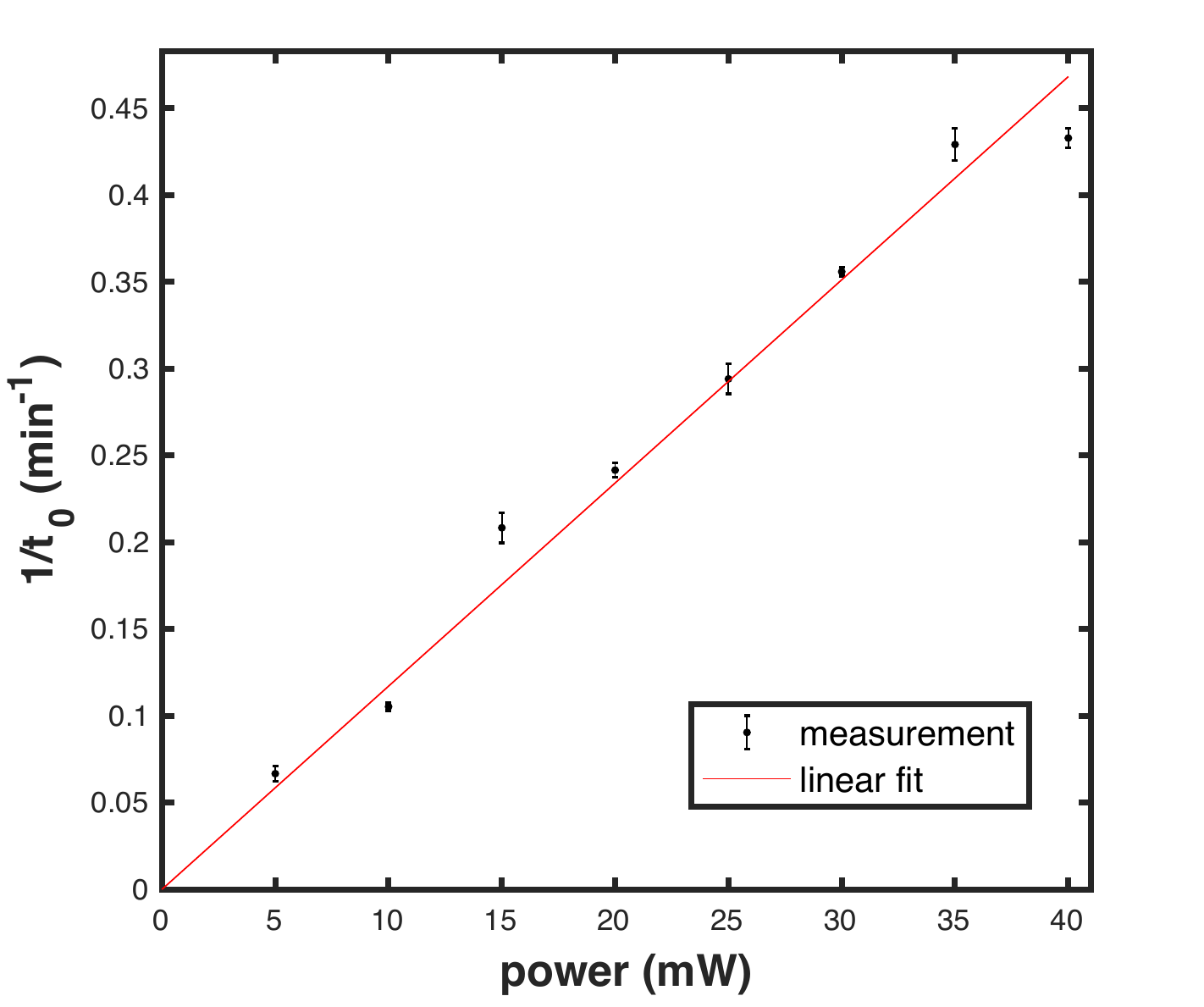}
\caption{Effect of laser power on the rate of change of the speckle pattern. The inverse of the decay time is shown as a function of the input power. The centre and span of the error bars respectively give the mean and standard deviation of $1/t_0$ in the duration of each measurement. We find an approximately linear relation with a proportionality constant of 0.012 $\text{min}^{-1}$/mW, which supports the heating-related origin of the effect. }
\label{fig:heating}
\end{figure}

In order to further confirm the hypothesis that the slow change in the speckle is due to heating from the laser, we measure $t_0$ for different powers of the input beam. Assuming again that the heat remains stored in the sphere's material, we expect the inverse of $t_0$ to be proportional to the input power. For each value of input power, we record the speckle patterns for 5 minutes, and extract $t_0$ by fitting with a Lorentzian profile. The result is shown in Fig. \ref{fig:heating}, where we also display as error bars the standard deviation of $1/t_0$ during the 5 minutes of each measurement. We find an approximately linear relation reading $1/t_0=0.012 P$, with $t_0$ in minutes and $P$ in milliwatts. From this we can infer the rate of change of the radius to be 0.1 $\text{nm.min}^{-1}\text{mW}^{-1}$. We note in passing that, while presently undesired, in future one could consider exploiting these effects to measure small changes in the temperature of the medium.

\section{Summary and conclusion}
In summary, we proposed a route to optimise speckle-based measurements of refractive index. While intuition suggests that the correct strategy is to maximise the path length of light in the medium of interest, it is more important to maximise the \emph{width} of the path length distribution within the medium. In particular, we have demonstrated that an integrating sphere, in which light has a broad path length distribution, offers a simple yet sensitive probe of refractive index change of the medium it encloses.
We quantified the change in the speckle pattern using the similarity (\ref{eq:correl}), analytically demonstrated that this takes a simple Lorentzian form as a function of refractive index change (\ref{eq:lorentzian_refrac}), and verified it experimentally. We gave a general expression for the HWHM of the similarity curve, and found that it depends mainly on the radius and surface reflectivity of the sphere, which paves the way for possible optimisations. In our setup, we found the HWHM to be $6.5\times10^{-7}$.

We exploited this high sensitivity to measure small refractive index variations of amplitude $4.5\times10^{-9}$ with an uncertainty of $7\times10^{-10}$. Our method allows a level of uncertainty comparable to current state of the art techniques, but with a significantly simpler implementation. On the other hand, it allows the measurement of variations (instead of absolute values) in refractive index and requires some care regarding heating effects due to the input laser light. We investigated this heating effect and found that 10 mW of laser power induces an increase in the sphere's radius of 1.1 nm every minute. This however can be compensated by applying either an appropriate volume or wavelength change. 

Importantly, the measurements presented here are three orders of magnitude more highly-resolved than previous implementations based on laser speckle. Developments to the existing apparatus to improve both the temperature stability and the resolution will be guided by the progress made in state-of-the-art Fabry-Perot systems \cite{Silander20}. In particular, the choice of material for the integrating sphere, plus the addition of active cooling, will significantly improve the thermal stability. Moreover, the use of shorter laser wavelength, larger image arrays and especially higher-reflectivity coating inside the sphere offer significant opportunities to measure even smaller refractive index changes. In the context of wavelength measurements, alternative forms of speckle analysis such as principal component analysis \cite{bruce19} and deep learning \cite{gupta19} have been shown to improve resolution and could also be applied to speckle refractometry.

\section*{Acknowledgements}
This work was supported by funding from the Leverhulme Trust (RPG-2017-197) and the UK Engineering and Physical Sciences Research Council (EP/P030017/1). The data that support the findings of this study will be openly available via the University of St Andrews Open Data Repository.

\bibliography{sample}
\end{document}